# Physical conditions in three high-$z$ H$_2$-bearing DLAs: implications for grain size

Katherine Rawlins[1], Gargi Shaw[1]† & Raghunathan Srianand[2]

[1]UM-DAE Centre for Excellence in Basic Sciences, University of Mumbai, Santa Cruz East, Mumbai 400 098, India
[2]IUCAA, Post Bag 4, Ganeshkhind, Pune University Campus, Pune 411 007, India

29 March 2016

**ABSTRACT**

We present results of our numerical simulation of three H$_2$-bearing damped Lyman $\alpha$ absorbers (DLAs) in the redshift interval ~ 2–3. The systems we have modelled are the DLAs at $z_{abs}$ = 2.3377 towards the quasar LBQS 1232+0815, at $z_{abs}$ = 2.41837 towards SDSS J143912.04+111740.5 and at $z_{abs}$ = 2.6265 towards FBQS J081240.6+320808. We have used the spectral synthesis code CLOUDY to simulate the physical environment of these DLAs, and constrain the density, radiation field, geometry and dust-grain properties of the DLAs self-consistently based on the observed column densities of various atomic and molecular species such as H I, fine structure lines of C I and the rotational level population of H$_2$. In our models, we explore the effect of grain size distribution on the predicted column densities of different species. Within the allowed uncertainties in the inferred dust-to-gas ratio, both models with standard ISM grains and smaller-sized grains reproduce the observations equally well. Improved constraints on dust-to-gas ratio and line-of-sight extinction are important for probing the grain size distribution in high-$z$ DLAs. We find the H$_2$-bearing clouds to have line-of-sight thickness in the range 1-6 pc, consistent with what has been found using partial coverage or 21-cm observations in some high-$z$ DLAs.

## 1 INTRODUCTION

Damped Lyman $\alpha$ absorbers (DLAs) are neutral gas clouds characterized by their high H I column densities, $N$ (H I) $\geq 10^{20.3}$ cm$^{-2}$ (Wolfe, Gawiser & Prochaska 2005). They trace bulk of the neutral hydrogen at high redshift (Noterdaeme et al. 2009, 2012) and are considered to be the progenitors of present-day disc galaxies (Wolfe et al. 1995). DLAs are detected in absorption along the line-of-sight to luminous high-redshift sources such as quasars or gamma-ray bursts. The mean DLA metallicity at $z_{abs}$ ~ 2-3 is ~1/15 Z$_\odot$, and the typical dust-to-gas ratio is ~ 1/30 of that in the Milky Way (Pettini et al. 1997). The physical conditions in DLAs vary widely from diffuse warm (n ~ 0.6 cm$^{-3}$, T ~ 5000 K) to the cold neutral medium (n ~ 30 cm$^{-3}$, T ~ 100 K) (Srianand et al. 2005; Draine 2011).

Physical conditions in the cold neutral medium are suitable for molecule formation as the interior of these regions are shielded enough to prevent photodissociation by far ultra-violet (FUV) radiation. H$_2$ is the most abundant molecule in DLAs, having been detected in ~ 10–15% of all DLA systems (Ledoux, Petitjean & Srianand 2003; Noterdaeme et al. 2008a). The H$_2$ lines can be observed through Lyman and Werner band absorption (11.2–13.6 eV) only at redshift $\geq$ 1.8 when the ultraviolet transitions redshift to the optical regime. Besides H$_2$, other molecules that have been detected in high-redshift DLAs are HD and CO (Varshalovich et al. 2001; Srianand et al. 2008). Cold gas that harbours molecules is also likely to be associated with star formation. In a thermally stable medium, heating is balanced by cooling. Radiative cooling in DLAs can be obtained from C II$^*$ column density measured through absorption lines. This cooling rate cannot be explained considering only the extragalactic background radiation contributed by galaxies and quasars (Wolfe, Prochaska & Gawiser 2003; Dutta et al. 2014). DLAs are thus, considered to be related to sites of on-going star formation. The typical star formation rate in high-$z$ DLAs is known to be low, $\leq$ 1–10 M$_\odot$ yr$^{-1}$ (Wolfe & Chen 2006; Fumagalli et al. 2010; Rahmani et al. 2010).

† Email: gargishaw@gmail.com

Observations of the rotational levels of H$_2$ along with the fine structure levels of C I, can be used to constrain models of cold gas in DLAs quite well (Srianand et al. 2005; Jorgenson, Wolfe & Prochaska 2010; Noterdaeme et al. 2015). The conventional notation for the various levels is H$_2$ ($v$, $J$), where $v$ is the vibrational level and $J$ is the rotational level. The level population can be used to constrain the particle density, radiation field and temperature of the gas. H$_2$ formation occurs both in the gas phase ($H^- + H \rightarrow H_2 + e^-$, $H_2^+ + H \rightarrow H_2 + H^+$) and on the surface of dust grains (Tielens & Hollenbach 1985; Tielens 2005; Shaw et al. 2005 and the references listed there; Draine 2011). However, the grain surface route is the dominant formation mechanism for H$_2$ in the metal-enriched environment. Hence, numerical simulations of H$_2$-bearing DLAs can be an effective tool to probe grain properties of these systems. C I which has an ionization potential 11.2 eV, is known to trace H$_2$ (Srianand et al. 2005). The fine structure levels $^3P_0$, $^3P_1$ and $^3P_2$, which we denote as C I*, C I** and C I*** respectively, also determine gas density through their mutual ratios.

Mathis, Rumpl & Nordsieck (1977), hereafter MRN, in their seminal work on size distribution of interstellar grains, used interstellar dust comprising of spherical particles of different materials such as graphite, silicon carbide, enstatite, olivine, iron and magnetite. They successfully fit the interstellar extinction from infrared to ultraviolet wavelengths using a size range of about 0.005 to 1 μm following a power-law size distribution with the exponent lying between -3.3 to -3.6. They use a variety of mixtures of the above-mentioned materials to fit the extinction and find that graphite is a necessary component which can be used in combination with any of the other particles. In this paper, the standard MRN grain size distribution is represented as $dn/da \propto a^{-3.5}$, where $n$ is the number of grains per unit volume and $a$ is the dust grain radius with a range of 0.005 to 0.250 μm. It is important to study grain properties such as size distribution at high-redshift as this has relevance for not just molecule formation but also galaxy evolution (Asano et al. 2013a, 2013b). Small carbonaceous polycyclic aromatic hydrocarbon molecules (PAHs) have also been detected in the ISM of star-forming galaxies through their infrared emission features (Smith et al. 2007). However, PAHs have not yet been observed in DLAs.

More than 25 detections of H$_2$ have been reported in DLAs and strong sub-DLAs along quasar sightlines for $z_{abs} > 1.8$ and $N$(H I) $\geq 10^{20.0}$ cm$^{-2}$ (Ledoux et al. 2003; Noterdaeme et al. 2008a; Albornoz-Vasquez et al. 2014 and references therein; Bagdonaite et al. 2014; Balashev et al. 2015; Klimenko et al. 2015; Noterdaeme et al. 2015). C I is seen in most of these DLAs. We selected 3 DLAs with good column density measurements of H$_2$ in various rotational levels of the ground vibrational state, and fine structure levels of C I, so as to maximize the observational constraints for our study. We have modelled these systems in detail using the spectral simulation code CLOUDY to understand the physical conditions in high-redshift DLAs. The 3 DLAs for which we study the physical environment are the systems at: $z_{abs}$ = 2.3377 towards LBQS 1232+0815 (Srianand, Petitjean & Ledoux 2000; Varshalovich et al. 2001; Srianand et al. 2005; Ivanchik et al. 2010; Balashev et al. 2011; Zafar et al. 2014), $z_{abs}$ = 2.41837 towards SDSS J143912.04+111740.5 (Noterdaeme et al. 2008b; Srianand et al. 2008), and $z_{abs}$ = 2.6265 towards FBQS J081240.6+320808 (Prochaska, Howk & Wolfe 2003; Jorgenson et al. 2009; Tumlinson et al. 2010; Jorgenson et al. 2010). Typically, DLAs may constitute absorption arising from more than one cloud along the line-of-sight. We are interested in modelling the molecular cloud, and provide details of the component/s that we have modelled in the respective sections.

This paper is organized in the following manner: In section 2, we describe our calculations for the 3 DLAs. Our results and conclusions are presented in sections 3 and 4, respectively.

## 2 CALCULATIONS

We have used the spectral synthesis code CLOUDY version c13.03 (Ferland et al. 1998; Ferland et al. 2013) to model these DLAs. This code is based on a self-consistent calculation of thermal, ionization, and chemical balance of gas and dust exposed to a source of radiation, and can be applied to different astrophysical systems. CLOUDY requires the spectral shape and intensity of the radiation field, chemical composition and density of the gas, and also geometry of the system as input. As radiation passes through the gas and dust in the cloud, the system is subject to various radiative and non-radiative processes which determine its physical and chemical state. CLOUDY divides the entire system into many zones and performs these calculations for each zone, which enables us to study the spatial variation of various quantities. It then predicts the spectrum along with column densities of various atomic and molecular species. Details of the chemical network, micro-physics of the H$_2$ molecule and grain physics are described in Abel et al. (2005), Shaw et al. (2005, 2006, 2008) and van Hoof et al. (2004), respectively.

The H$_2$ model in CLOUDY described in Shaw et al. (2005) includes all the published ro-vibrational levels within the ground and the lowest 6 electronically excited states. The formation rate of H$_2$ on grains depends on various parameters such as dust-grain type, dust and gas temperature, hydrogen density and total available grain surface area. It can proceed through either the Langmuir-Hinshelwood (LH) mechanism or the Eley-Rideal (ER) mechanism. In the LH mechanism which dominates at grain temperatures ≤ 100 K, two H atoms are physisorbed and migrate towards each other through quantum mechanical tunnelling or thermal hopping. The ER mechanism involves recombination of a chemisorbed H atom with an H atom from the gas phase (Cazaux & Tielens 2002). Both these processes have been considered in CLOUDY. Graphite and silicate grains are considered in the treatment of the H$_2$ model. The vibrational levels of a newly formed H$_2$ molecule on grain surface are distributed as per Takahashi & Uehara (2001), while the rotational level distribution is based on Takahashi (2001). Ortho-para conversion on grain surface is also considered. The conversion from ortho to para state on grain surface at low temperatures is treated according to Le Bourlot (2000).

The grain model in CLOUDY uses a Mie code based on Hansen & Travis (1974) to calculate absorption & scattering opacities for homogenous, spherical dust grains of different types. Grains can be carbonaceous, silicates or PAHs. They can either be single-sized or made to follow an arbitrary size distribution. The grain sizes can be resolved into a user-specified number of bins. We use graphites and silicates following an MRN or MRN-like power-law size distribution resolved in 10 size bins for our calculations. Grain properties such as temperature, charge distribution and opacities are determined by CLOUDY for each size bin separately since they depend on grain size. The grain charge distribution in each bin is resolved into discrete charge states. We use the default two charge states. The populations of the charge states are determined by self-consistently solving for the grain ionization-recombination balance (van Hoof et al. 2004). Photoelectric effect, quantum heating and various collisional processes are treated in detail, taking into account both grain type and size. Various grain surface reactions which are important for molecule formation are also considered.

Warm and cold neutral phases are known to co-exist in pressure equilibrium in the Galactic ISM (Field, Goldsmith & Habing 1969; Wolfire et al. 1995, 2003).We model the above-mentioned three DLAs as constant-pressure gas clouds with plane parallel geometry. The various components of pressure considered in our models are pressure arising from radiation, turbulence, and thermal motion of the gas. The sum of these components remains the same at every point in the cloud, though the individual components may vary with depth. Since we had no a priori knowledge of the temperature and density profile within the cloud, we consider constant pressure in our models. We find the temperature in our models to be almost constant throughout the cloud. Hence, in the absence of components like magnetic and hydrodynamic pressure, our constant pressure DLA models are equivalent to constant density models. Radiation is incident on these constant-pressure DLA clouds from both sides with equal intensity. This includes the radiation field emerging from the local star formation. Earlier, Srianand et al. (arXiv: 0506556) have used such local star formation for their DLA. Along with the radiation from the local star formation, our calculations also include the Haardt & Madau metagalactic background (Haardt & Madau 2001) and cosmic microwave background (CMB) at appropriate redshift. For the DLA at $z_{abs}$ = 2.6265, the UV interstellar radiation from the hot, massive O/B type stars is simulated through a blackbody radiation field with temperature 40,000 K. However, for the DLAs at $z_{abs}$ = 2.3377 & 2.41837, we have to use an X-ray radiation field. We discuss the details of this radiation field in sections 2.2 and 3.2. We also study the effect of purely one-sided radiation, which we discuss in section 3.3. Our models take into account the ionization caused by cosmic rays. We use observed elemental abundances wherever available, and include silicate and graphite dust grains of equal volume in our models. The dust grain sizes follow a power-law distribution. The dust-to-gas ratio, which is the ratio of the dust content of the system to that of the Milky Way, is calculated from observed column densities using the relation from Wolfe et al. (2003)

$$\kappa = 10^{[X/H]_{int}}\left(10^{[Y/X]_{int}} - 10^{[Y/X]_{gas}}\right) \qquad (1)$$

where X is a volatile element that is undepleted on dust grains such as S, Zn or Si, and Y is a refractory element such as Fe, Cr or Ni. $[X/H]_{int}$ is the intrinsic (undepleted) abundance of the volatile element. $[Y/X]_{int}$ and $[Y/X]_{gas}$ are the intrinsic and gas phase ratio of the refractory to volatile element, respectively. All these quantities are specified with respect to solar abundances. We refer to the solar abundances of Grevesse et al. (2010). We assume $[X/H]_{int} = [X/H]_{gas}$ since X is undepleted, and $[Y/X]_{int} = [Y/X]_\odot$. The calculated dust abundance is then used as a constraint for our models. All calculations stop at a value of total $H_2$ column density which is within the range provided by observation. We constrain our models through the column densities of the other observed species such as H I, rotational level population of $H_2$ and fine structure levels of C I. The physical parameters that we have varied in our calculations are parameters corresponding to the radiation field of the local star formation, hydrogen density, metal abundances, grain sizes and abundance, micro-turbulence and H I cosmic ray ionization rate. We vary the different input parameters individually, and study the effect on the predicted column densities. We then try to converge to the observed column densities of various species and determine the best-fitting model.

## 2.1. DLA at $z_{abs}$ = 2.6265 towards the quasar FBQS J081240.6+320808

This system was observed using the High Resolution Echelle Spectrograph (HIRES) on the Keck I Telescope (Prochaska et al. 2003; Jorgenson et al. 2009; Tumlinson et al. 2010; Jorgenson et al. 2010). There are 2 DLAs along this sightline − at $z_{abs}$ = 2.066780, and a multi-component DLA at $z_{abs}$ = 2.626, which is the system of our interest. The absorption from this DLA is spread over three components, at $z_{abs}$ = 2.625890, 2.626310 and 2.626491, with a possible fourth broad component at $z_{abs}$ = 2.626447. Here, we model component 3, which has the highest $N$ (H I) and contains most of the $H_2$.

We constrain our model parameters through the fine structure lines of C I and the rotational level population of $H_2$. We consider the C I fine structure level column densities reported in table 2 of Jorgenson et al. (2009). For the $H_2$ level population, Jorgenson et al. (2010) provide 2 models – model a, with the Doppler parameter of $H_2$ tied to C I, and model b, with the Doppler parameter derived from the (0, 3) level of $H_2$. The column densities of the levels $H_2$ (0, 0), (0, 1) and (0, 4) in the two models are within a difference of 0.02 dex, while for the levels $H_2$ (0, 2) and (0, 3) the difference is over 0.2 dex. We therefore, consider the lower and upper limits for each level by taking both models into account, and try to produce column densities within this combined range. We refer to the column densities of H I, Cr II and Zn II listed in table 7 of Jorgenson et al. (2010). The value of $N$ (H I) provided in this table has been scaled to trace $N$ (Zn II) in this particular component of the DLA. The total $N$ (H I) for the entire DLA is much higher, at 21.35 (log-scale) (table 3 of Jorgenson et al. 2010). Jorgenson et al. (2010) obtain $N$ (C II*) by two methods – Voigt profile fitting which yields column densities for individual components of the DLA, and the apparent optical depth method (AODM – Savage & Sembach 1991) which is used to derive the total column density over the entire DLA profile. The values in log-scale are 15.14 (from profile fit for component 3) and 14.30 (AODM over the entire DLA) respectively.

We model this DLA as a constant-pressure cloud with radiation impinging both sides. Here, we use a blackbody radiation field of temperature 40,000 K to simulate the UV radiation from the local star formation. The intensity of this radiation field is 0.3 times the Habing field, where the Habing field is the typical interstellar radiation density between 912 and 2400 Å (Habing 1968). The density of the gas is ~89 cm$^{-3}$, which is within the range of 40–200 cm$^{-3}$ suggested in Jorgenson et al. (2009). We also include cosmic rays in our model with ionization rate of neutral hydrogen 2 × 10$^{-17}$ s$^{-1}$. This is 1 dex lower than the Galactic background rate of 2 × 10$^{-16}$ s$^{-1}$ (Indriolo et al. 2007). We constrain the cosmic ray ionization rate mainly through its effect on the population of higher rotational levels of $H_2$, though H I and the C I fine structure levels are also sensitive to it. C II* is also enhanced by increased cosmic ray ionization. We adopt solar metallicity for this DLA as reported by Tumlinson et al. (2010) and separately set abundances for Cr and Zn according to the observations in table 7 of Jorgenson et al. (2010). The O abundance in our model is taken from Prochaska et al. (2003). We scale C, Mg and Si abundances to match observed column densities of C I, Mg I and Si I, though there are no observational constraints on their elemental abundances. We find that our predicted $N$ (CO) is less than 12

(log-scale) for the C abundance in our model, which agrees with the non-detection of CO for this DLA. We use $N$ (Si I) to constrain the Si abundance, since $N$ (Si II) is not available for this system. However, Si I has ionization potential 8.15 eV, and most of the silicon in DLAs is typically known to be present in Si II. Hence, Si I may not be efficient in constraining the Si abundance. [Si/H] in our model is 2 dex lower than solar, and also much lower than [Zn/H]. Zn abundance is well constrained with [Zn/H] in our model (-0.47) matching with the value inferred from observation (-0.58) by Jorgenson et al. (2010). It thus, appears that the [Si/H] obtained from $N$ (Si I) is not a true indicator of Si abundance in this DLA.

Jorgenson et al. (2010) calculate the dust-to-gas ratio for this component of the DLA to be 0.16. However, using the observed $N$ (Cr II) and $N$ (Zn II) and solar abundances from Grevesse et al. (2010) in Eq. (1), we derive dust-to-gas ratio 0.29. Depending on which dust abundance value is used as constraint, we find that the DLA can be modelled in two ways. Accordingly, we present Model 1 and Model 2 for this DLA. Model 1 has dust abundance 0.24 ISM, which is close to our calculated dust abundance. The grains are of ISM size (size limits 0.005–0.250 μm) and follow the MRN distribution. Model 2 uses smaller grains of size 0.5 times the ISM size (size limits 0.0025–0.125 μm) distributed according to the same power law as the MRN distribution. This model requires a dust abundance 0.12 ISM, which agrees with the value calculated by Jorgenson et al. (2010). All other input parameters except for these grain properties are the same between these two models. The input parameters of both models are listed in Table 1.

Both the models reproduce the various observed column densities satisfactorily within observational error. Our model prediction for $N$ (H I) agrees within ~0.15 dex of the observed $N$ (H I) for this component of the DLA. Our prediction for $N$ (C II*) = 14.05, which is lower by 0.25 dex than the value derived through AODM over the entire DLA profile. This is acceptable since we model here only the H$_2$-bearing component, and the remaining contribution to $N$ (C II*) can be understood to arise from the other components associated with this DLA. The observed and predicted column densities of various species are shown in Table 2.

In Fig. 1, we show the effect of different dust abundances and dust grain sizes on the two models we have presented above. In the left panel, we plot the ratio of the model-to-observed column density for various species for two models having dust abundance 0.12 ISM and 0.24 ISM respectively. We retain the ISM grain size (0.005-0.250 μm) and MRN power-law size distribution for the grains. All other parameters are held fixed between the two models. In the right panel, we consider a fixed dust abundance of 0.12 ISM and show the changes produced in the column densities of various species merely by changing the grain sizes. We plot the ratio of the model-to-observed column density for various species when the grain sizes are 0.2 ISM (0.001–0.050 μm), 0.5 ISM (0.0025–0.125 μm) and ISM (0.005–0.250 μm). All 3 models follow the same power law as the MRN size distribution. Again, all other parameters in the models are held fixed. It is clear that higher abundance of ISM-sized grains is necessary to reproduce the observed column densities, and also that the same model can be replicated with lower dust abundance by including smaller grains. We further discuss the implications of grain size and abundance in section 3.1.

The total extent of the cloud is 2.95 pc. The predicted gas pressure from our model varies in the range 4150–4660 K cm$^{-3}$, which matches with observation. Jorgenson et al. (2009) calculate a gas pressure $P/k$ in the range 2500–16,000 K cm$^{-3}$. The electron temperature reaches 48 K in the interior of the DLA, while the kinetic temperature derived from both H$_2$ models of Jorgenson et al. (2010) lies in the range $T_{01}$= 44–48 K. All these values are from the predictions of Model 2. Model 1 yields similar values. Fig. 2 shows the variation of abundances, temperature, heating and cooling mechanisms with depth from one of the two illuminated faces of the cloud. The other half of the cloud will have a symmetrical profile for each of these quantities. This DLA is predominantly heated by the grain photoelectric mechanism which is responsible for ~ 85 percent of the heating fraction, while cosmic rays and H$_2$ photodissociation contribute ~ 5-6 percent each. As can be seen in panel 3 of Fig. 2, the grain photoelectric mechanism and cosmic rays provide heating almost uniformly throughout the extent of the DLA. The contribution of H$_2$ photodissociation rises slightly deep into the cloud in the H$_2$-forming region at distance ~ $10^{17}$ cm, and drops again at ~ $10^{18}$ cm owing to H$_2$ self-shielding. The C II* 158 μm line emission is the main coolant, with a contribution in excess of 94 percent throughout the cloud. The O I* 69 μm line accounts for ~ 4 percent of the cooling. The heating and cooling fractions discussed here are with respect to Model 2, which uses smaller grains. Model 1 agrees to within 10 percent for all quantities. Since the trends are similar for both models, we have shown in Fig. 2, the results only from Model 2.

Further, we check our model predictions for other atomic and molecular species, so that they can be observed in future to validate our model. We predict high log column densities for the following species: OH and HCl at 12.53 and 13.64 respectively from Model 2 presented above. The predictions of both models agree to within 0.1 dex for all species.

**Table 1.** Input parameters for our CLOUDY models for the DLA at $z_{abs}$ = 2.6265 towards the quasar FBQS J081240.6+320808

| Parameter | Values from observation | Model 1 | Model 2 |
|---|---|---|---|
| Hydrogen density, $n_H$ | 40–200 cm$^{-3}$ [a,b] | 89–94 cm$^{-3}$ | 89–94 cm$^{-3}$ |
| Intensity of blackbody radiation | | $5 \times 10^{-4}$ ergs cm$^{-3}$ s$^{-1}$ | $5 \times 10^{-4}$ ergs cm$^{-3}$ s$^{-1}$ |
| [C/H] | | -1.25 | -1.25 |
| [O/H] | -0.54±0.1 [c] | -0.42 (-0.60) [d] | -0.42 (-0.60) [d] |
| [Mg/H] | | -1.52 [e] | -1.52 [e] |
| [Si/H] | | -2.00 [e] | -2.00 [e] |
| [Cr/H] | ≤ -2.89 [f] | -3.00 (-3.05) [g] | -3.00 (-3.05) [g] |
| [Zn/H] | -0.58 [f] | -0.47 (-0.58) [g] | -0.47 (-0.58) [g] |
| Metallicity (abundance for other species) | $Z_\odot$ [a] | $Z_\odot$ | $Z_\odot$ |
| Grain size range [h] | | 0.005–0.250 μm (ISM) | 0.0025–0.125 μm (0.5 ISM) |
| Dust-to-gas ratio, $\kappa$ | 0.16 [i] | 0.24 ISM | 0.12 ISM |
| Doppler parameter, $b$ | 1.19 km s$^{-1}$ [i] | 1.2 km s$^{-1}$ | 1.2 km s$^{-1}$ |
| Cosmic ray ionization rate, $\zeta_H$ | | $2 \times 10^{-17}$ s$^{-1}$ | $2 \times 10^{-17}$ s$^{-1}$ |

[a] From Jorgenson et al. (2009).
[b] Gas density is determined through the rate equations of the C I fine structure levels, assuming the gas temperature to be constrained by $T_{01}$, here 44–48 K.
[c] Prochaska et al. (2003).
[d] Abundances are with respect to the solar abundances in Grevesse et al. (2010), which is the solar standard used by CLOUDY. The abundance values in brackets are the values with respect to solar abundances in Savage & Sembach (1996).
[e] Abundances are with respect to the solar abundances in Grevesse et al. (2010), which is the solar standard used by CLOUDY.
[f] Values from table 7 of Jorgenson et al. (2010) with respect to the solar abundances in Morton (2003).
[g] Abundances are with respect to the solar abundances in Grevesse et al. (2010), which is the solar standard used by CLOUDY. The abundance values in brackets are the values with respect to solar abundances in Morton (2003).
[h] MRN or MRN-like power-law size distribution.
[i] From Jorgenson et al. (2010). Dust-to-gas ratio from table 7, and Doppler parameter from table 8 therein.

**Table 2.** Observed column densities and our CLOUDY model predictions for the DLA at $z_{abs}$ = 2.6265 towards the quasar FBQS J081240.6+320808. Model 1 uses ISM-sized grains with abundance 0.24 ISM (which agrees with our calculations), and Model 2 uses smaller-sized grains (0.5 times ISM) with abundance 0.12 ISM (which agrees with the value quoted by Jorgenson et al. (2010)).

| Species (X) | Observed column densities log N(X) cm$^{-2}$ | Model 1 predicted column densities log N(X) cm$^{-2}$ | Model 2 predicted column densities log N(X) cm$^{-2}$ |
|---|---|---|---|
| H I | 20.97[a] | 20.83 | 20.84 |
| H$_2$ | 19.88±0.20[b] | 19.85 | 19.85 |
| H$_2$ (0, 0) | 19.76–19.84[c] | 19.74 | 19.75 |
| H$_2$ (0, 1) | 19.10–19.18[c] | 19.21 | 19.17 |
| H$_2$ (0, 2) | 16.10–16.63[c] | 16.14 | 16.11 |
| H$_2$ (0, 3) | 14.81–15.16[c] | 15.09 | 15.09 |
| H$_2$ (0, 4) | 13.93–14.03[c] | 14.02 | 14.03 |
| C I* | 13.30±0.20[d] | 13.37 | 13.38 |
| C I** | 13.02±0.03[d] | 13.15 | 13.16 |
| C I*** | 12.47±0.05[d] | 12.34 | 12.33 |
| C II* | 14.30±0.01[e] | 14.07 | 14.05 |
| Mg I | 12.68[d] | 12.71 | 12.72 |
| Si I | 11.53±0.18[d] | 11.44 | 11.44 |
| Cr II | ≤ 11.71[a] | 11.55 | 11.56 |
| Zn II | 13.00[a] | 13.00 | 13.01 |

[a] Table 7 of Jorgenson et al. (2010).
[b] From Tumlinson et al. (2010).
[c] Table 8 of Jorgenson et al. (2010). We consider the combined range of column densities of models a and b for component 3, including errors.
[d] From Jorgenson et al. (2009). Column densities of C I fine structure lines from table 2 therein.
[e] Obtained by the apparent optical depth method. Refer text in section 2.1.

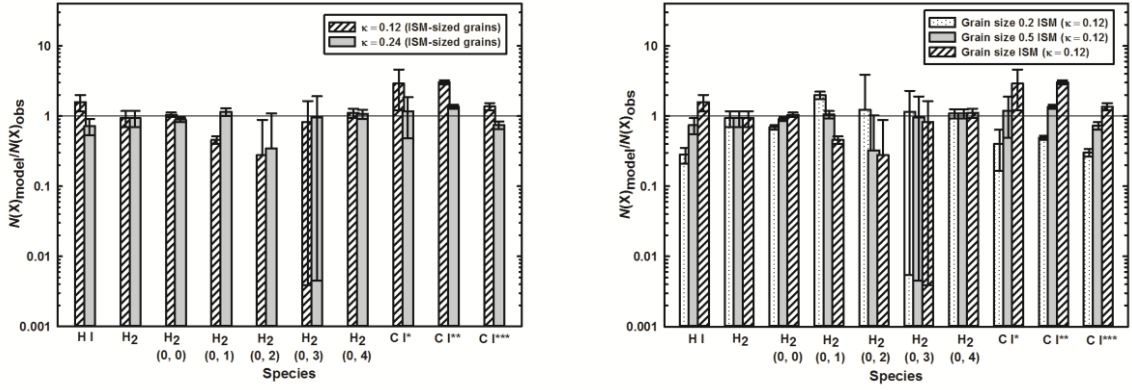

**Figure 1.** The plot shows the ratio of model-to-observed column densities for H I, $H_2$ and C I for the DLA at $z_{abs}$ = 2.6265 towards the quasar FBQS J081240.6+320808. In the left panel, the dust abundance is varied for ISM-sized grains while all other input parameters are held fixed. In the right panel, only the grain size is varied for a fixed dust abundance of 0.12 ISM to show its effect on the column densities of various species. The errors have been calculated by considering the ratio of our model predictions with the upper and lower limits of observed column density.

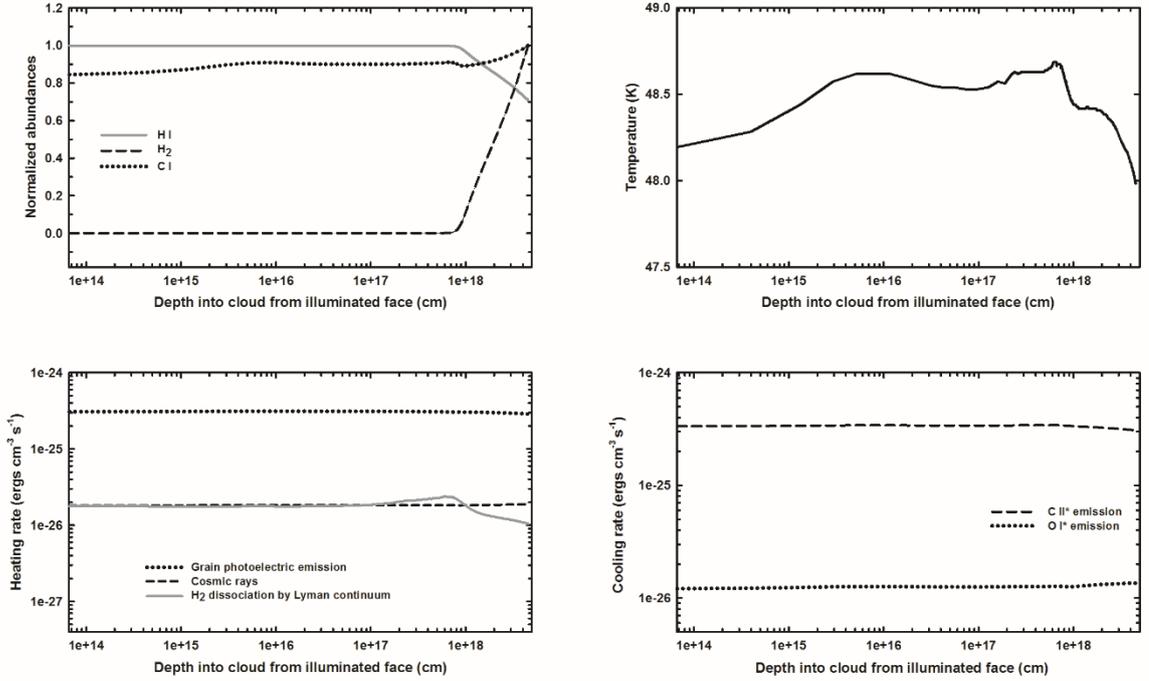

**Figure 2.** The physical conditions of the DLA at $z_{abs}$ = 2.6265 towards the quasar FBQS J081240.6+320808, as predicted by Model 2 are summarized in this figure. The upper left panel shows the abundance profile for H I, $H_2$ and C I; while the upper right panel shows the temperature variation through the DLA. The lower left and right panels show the variation of the heating caused due to grain photoelectric emission, cosmic rays and $H_2$ photodissociation; and the cooling due to C II* and O I* emission, respectively. Model 1 shows similar variation trends.

## 2.2. DLA at $z_{abs}$ = 2.41837 towards the quasar SDSS J143912.04+111740.5

Observations for this DLA were carried out at the Very Large Telescope (VLT), Chile using the Ultraviolet and Visual Echelle Spectrograph (UVES) (Srianand et al. 2008; Noterdaeme et al. 2008b). This was the first instance of detection of CO in a DLA. $H_2$ is seen in 6 components spread over only 50 km/s, and CO has been detected in 2 of these components. Srianand et al. (2008) derive component-wise column densities for CO and its lowest three rotational levels, but only total $H_2$ column density over the multiple components for the $H_2$ (0, 0) and $H_2$ (0, 1) levels on account of damping of the transitions. Since we are interested in studying $H_2$, we model the entire molecular cloud associated with this DLA. The fine structure levels of C I are also observed. The C II* λ1335.7 absorption line was not covered by the UVES observations of this DLA. Due to the presence of Lyman-α, β and γ absorption lines, component-wise $N$ (H I) has been determined by Srianand et al. (2008). The highest log column density of $N$ (H I) = 20.10, was found to be associated with the molecular component. We try to match this column density in our model.

This DLA is also modelled as a constant-pressure cloud with radiation incident on both faces. But here, we find that a blackbody radiation field of temperature 30,000-40,000 K does not succeed in reproducing the observed column densities satisfactorily. We try two different scenarios – first, where we model only the cold neutral medium by taking into account the attenuation of the incident radiation after it has passed through an absorbing medium, and second, where we include the surrounding ionized H II region (temperature ~ $10^4$ K) too. By considering only the cold phase of gas, we are unable to simultaneously produce the observed column density of CO, $H_2$ (0, 0) and $H_2$ (0, 1). Including the surrounding ionized region in our models does not reproduce the observed mutual ratios of the C I fine structure levels. The upper levels are excited by the harsh radiation to a greater extent than observed. We thus move on to a third scenario, employing a different type of radiation field. We incorporate a hard X-ray radiation field extending over the energy range 1-100 keV, and model only the cold neutral phase of the gas. This radiation field follows the spectral energy distribution described by Maloney, Hollenbach & Tielens (1996) and has negligible intensity outside the above-mentioned energy range. We use logarithmic ionization parameter of -5, where ionization parameter is the ratio of ionizing photon density to the hydrogen density, to reproduce all observed quantities satisfactory. This radiation field has UV intensity 0.4 $G_0$ (Tielens & Hollenbach 1985). Our model for this DLA is thus, similar to an X-ray Dominated Region (XDR). We discuss the XDR radiation field in detail in section 3.2.

For a given density and ionization parameter, the column densities of CO and the C I fine structure lines can be constrained simultaneously by varying the C and O abundances and the cosmic ray ionization rate. We therefore, scale C and O abundances, though there are no observational constraints on them. We also vary the ionization parameter of the radiation field, gas density and cosmic ray ionization rate to determine the best fit. The hydrogen density in this DLA is ~ 60 cm$^{-3}$, which is in agreement with the range of 45–62 cm$^{-3}$ based on the observed column density ratios of the C I fine structure lines in the strongest CO component (Srianand et al. 2008). Since $H_2$ is detected only in the lowest two rotational levels in this DLA, the column densities of CO and C I fine structure lines are the quantities that constrain the cosmic ray ionization rate. Our model for this DLA needs cosmic ray ionization rate 1.4 ×$10^{-15}$ s$^{-1}$, which is 7 times the background value attributed to the Galactic ISM. We find that this is required to increase the predicted CO column density to match observation. In accordance with observation, we apply solar elemental abundances to our model and separately scale the abundances of N, Si, S, Fe and Zn as per the observed data in table 1 of Noterdaeme et al. (2008b).

We use Eq. (1) to calculate the dust-to-gas ratio in this case also. This system has observed column densities of Fe II, Si II, S II and Zn II. Thus, we determine the dust-to-gas ratio by considering each of the undepleted species individually with Fe. They give vastly different estimates of the dust-to-gas ratio. [Fe/Si] yields dust abundance 0.11±0.04, while [Fe/S] and [Fe/Zn] yield the values 1.07±0.26 and 1.81±0.43 respectively. Since Si is known to be mildly refractory, we ignore the estimate obtained through [Fe/Si] (Prochaska & Wolfe 2002; Vladilo et al. 2011). S and Zn are both undepleted on grain surface, yet the dust-to-gas ratio estimates from them both show significant difference. The dust abundance calculated using [Fe/Zn] implies the DLA dust content is almost twice that of the local ISM, which is a very high value. The vast difference in the calculated dust-to-gas ratio affects the constraints we impose on our model. Here again, we develop two models like for the DLA at $z_{abs}$ = 2.6265. Model 1 uses ISM-sized grains, while Model 2 uses grains 0.5 times the ISM size. In Model 1, we find that we require dust-to-gas ratio 1.9 ISM, which is in agreement with the high dust-to-gas ratio derived from [Fe/Zn]. However, Model 2 requires dust abundance similar to the ISM, which matches the abundance calculated from [Fe/S]. All input parameters except the above-mentioned grain properties are the same for Models 1 and 2. The input parameters of the two models are listed in Table 3. We are able to predict all the observed column densities well. The observed and predicted column densities of both models are shown in Table 4. Since we are able to replicate the physical conditions within the DLA by coupling a change in dust abundance by factor ~2, with factor 2 changes in the grain radii, our models seem to indicate the need for increased grain surface area. Robust conclusions about the size of the dust grains cannot be made unless the dust-to-gas ratio is constrained accurately.

In Fig. 3, the left panel shows the effect of using different abundances of ISM-sized grains on column densities of different species. The right panel shows the predictions of three models utilizing different grain sizes (0.2 ISM, 0.5 ISM and ISM size), where the dust-to-gas ratio is fixed at the ISM value. All models follow the power law of the MRN size distribution. Other parameters apart from the mentioned grain properties are held fixed in the models.

This cloud is smaller than the previous one, extending to a depth of only 1.02 pc. The predicted gas pressure from our model for this DLA is in the range 5620–5660 K cm$^{-3}$. The sightline through this DLA traverses deep through the molecular region, far from the H I/$H_2$ transition region. The shielding provided by $H_2$ enhances the formation of CO. The electron temperature reaches 99 K in the interior of the DLA. This agrees with the kinetic temperature derived from $H_2$, $T_{01} = 105^{+42}_{-32}$ K (Srianand et al. 2008). All these values are from the predictions of Model 2. Model 1 yields similar values. The variation of the main heating and cooling processes with depth is shown in Fig. 4, along with the temperature and abundance profiles. The plots again trace only one half of the cloud, with the other half producing a symmetrical profile. Since the modelled region is mostly molecular, the heating is mainly due to cosmic rays, which causes ~ 50 percent of the heating. Grain photoelectric heating is the second major contributing process with heating fraction ~ 35 percent. Cooling occurs through the C II* 158 μm and O I* 69 μm lines, with C II* being the dominant coolant, accounting for ~ 80 percent of the cooling. O I* is responsible for ~ 15 percent of the total cooling. Grain photoelectric heating and cosmic ray heating have almost uniform effect throughout, while heating due to photoionization of hydrogen peaks at a depth ~ $10^{16}$ cm and then falls away. $H_2$ photodissociation becomes a significant contributor to the gas heating around the H I/ $H_2$ transition region where $H_2$ is photodissociated into H. Collisional deexcitation of $H_2$ acquires significance in the molecular region beyond $10^{16}$ cm, where it is the main influence on temperature as can be seen from the rise and plateau in the temperature profile. The heating and cooling fractions discussed here are with respect to Model 2, which uses smaller grains. Model 1 agrees with Model 2 within 10 percent for most quantities. Grain photoelectric heating is higher in case of Model 1, and agrees with Model 2 to ~30 percent. In the molecular region, where heating due to collisional deexcitatation of vibrationally-excited $H_2$ is important, Model 1 shows 15-20 percent increased heating over Model 2. However, all quantities show similar variation trends across the cloud. We have shown in Fig. 4, the results from Model 2.

To study the effect of CMB on CO level populations, we compute our model for this DLA with CMB redshift set to zero keeping other parameters same. At $z = 0$, CMB temperature is 2.7 K and at z = 2.41837, it is 9.2 K. The two models show a marked difference in the mutual ratios of the CO levels. In the T = 2.7 K case, most of the CO is produced in the (0, 0) level, while in our actual model for $z = 2.41837$ (or with T = 9.2 K), most of the CO lies in the (0, 1) level. This shows that CMB does play a vital role in deciding the CO population distribution. The pumping at high-$z$ occurs because the CMB temperature at those redshifts is comparable to the energy gaps between the lowest few rotational levels of CO. Collisions do not play an important role here as the critical density for these levels is much higher than the density seen in the DLA. Our calculations thus confirm the reasoning of Srianand et al. (2008) based on calculations using the statistical equilibrium radiative transfer code RADEX (van der Tak et al. 2007).

For this DLA, Model 2 predicts high C II*, OH, OH$^+$ and H$_2$O log column densities at 14.39, 14.13, 13.13 and 12.79 respectively. Predictions from Model 1 agree to within 0.1 dex for all species. Future observations of this sightline with improved detection abilities may be able to detect transitions of these species.

**Table 3.** Input parameters for our CLOUDY models for the DLA at $z_{abs}$ = 2.41837 towards the quasar SDSS J143912.04+111740.5

| Parameter | Values from observation | Model 1 | Model 2 |
|---|---|---|---|
| Hydrogen density, $n_H$ | 45–62 cm$^{-3}$ [a, b] | 60-61 cm$^{-3}$ | 60-61 cm$^{-3}$ |
| Logarithmic ionization parameter | | -5 | -5 |
| [C/H] | | -0.57 | -0.57 |
| [N/H] | $\geq$-0.34 [c] | -0.22 (-0.34) [d] | -0.22 (-0.34) [d] |
| [O/H] | | 0.15 | 0.15 |
| [Si/H] | -0.86±0.11 [c] | -0.92 (-0.97) [d] | -0.92 (-0.97) [d] |
| [S/H] | -0.03±0.12 [c] | -0.07 (-0.15) [d] | -0.07 (-0.15) [d] |
| [Fe/H] | -1.32±0.11 [c] | -1.43 (-1.43) [d] | -1.43 (-1.43) [d] |
| [Zn/H] | +0.16±0.11 [c] | +0.11 (+0.05) [d] | +0.11 (+0.05) [d] |
| Metallicity (abundance for other species) | $Z_\odot$ [e] | $Z_\odot$ | $Z_\odot$ |
| Grain size range [f] | | 0.005–0.250 μm (ISM) | 0.0025–0.125 μm (0.5 ISM) |
| Dust-to-gas ratio, κ | 0.11±0.04 (From [Fe/Si]) 1.07±0.26 (From [Fe/S]) 1.81±0.43 (From Fe/Zn]) | 1.9 ISM | ISM |
| Doppler parameter, b | 1.5 km s$^{-1}$ [a] | 1.5 km s$^{-1}$ | 1.5 km s$^{-1}$ |
| Cosmic ray ionization rate, $\zeta_H$ | | 1.4 × 10$^{-15}$ s$^{-1}$ | 1.4 × 10$^{-15}$ s$^{-1}$ |

[a] Srianand et al. (2008).
[b] Gas density is determined through the rate equations of the C I fine structure levels, assuming the gas temperature to be constrained by $T_{01}$, here $105^{+42}_{-32}$ K.
[c] Values from table 1 of Noterdaeme et al. (2008b) with respect to the solar abundances in Morton (2003).
[d] Abundances are with respect to the solar abundances in Grevesse et al. (2010), which is the solar standard used by CLOUDY. The abundance values in brackets are the values with respect to solar abundances in Morton (2003).
[e] Solar metallicity DLA as per Noterdaeme et al. (2008b)
[f] MRN or MRN-like power-law size distribution.

**Table 4.** Observed column densities and our CLOUDY model predictions for the DLA at $z_{abs}$ = 2.41837 towards the quasar SDSS J143912.04+111740.5. Model 1 uses ISM-sized grains with abundance 1.9 ISM (which agrees with the dust-to-gas ratio calculated from [Fe/Zn]), and Model 2 uses smaller-sized grains (0.5 times ISM) with ISM abundance (which agrees with the dust-to-gas ratio calculated from [Fe/S]).

| Species (X) | Observed column densities log N(X) cm$^{-2}$ | Model 1 predicted column densities log N(X) cm$^{-2}$ | Model 2 predicted column densities log N(X) cm$^{-2}$ |
|---|---|---|---|
| H I | 20.10±0.10 [a] | 20.14 | 20.15 |
| H$_2$ | 19.38±0.10 [a] | 19.33 | 19.38 |
| H$_2$ (0, 0) | 18.90±0.10 [a] | 18.88 | 18.95 |
| H$_2$ (0, 1) | 19.18±0.10 [a] | 19.13 | 19.17 |
| CO | 13.89±0.02 [a] | 13.87 | 13.88 |
| CO (0, 0) | 13.38±0.06 [a] | 13.26 | 13.28 |
| CO (0, 1) | 13.58±0.05 [a] | 13.51 | 13.53 |
| CO (0, 2) | 13.18±0.06 [a] | 13.24 | 13.26 |
| C I* | 14.57±0.01 [b] | 14.53 | 14.54 |
| C I** | 14.28±0.02 [b] | 14.30 | 14.30 |
| C I*** | 13.44±0.03 [b] | 13.53 | 13.52 |
| N I | $\geq$ 15.71 [c] | 15.87 | 15.89 |
| Si II | 14.80±0.04 [c] | 14.85 | 14.86 |
| S II | 15.27±0.06 [c] | 15.31 | 15.33 |
| Fe II | 14.28±0.05 [c] | 14.33 | 14.34 |
| Zn II | 12.93±0.04 [c] | 12.89 | 12.91 |

[a] From Srianand et al. (2008).
[b] From C I profile fit. Column densities for only the strongest CO component are mentioned in Srianand et al. (2008).
[c] Table 1 of Noterdaeme et al. (2008b).

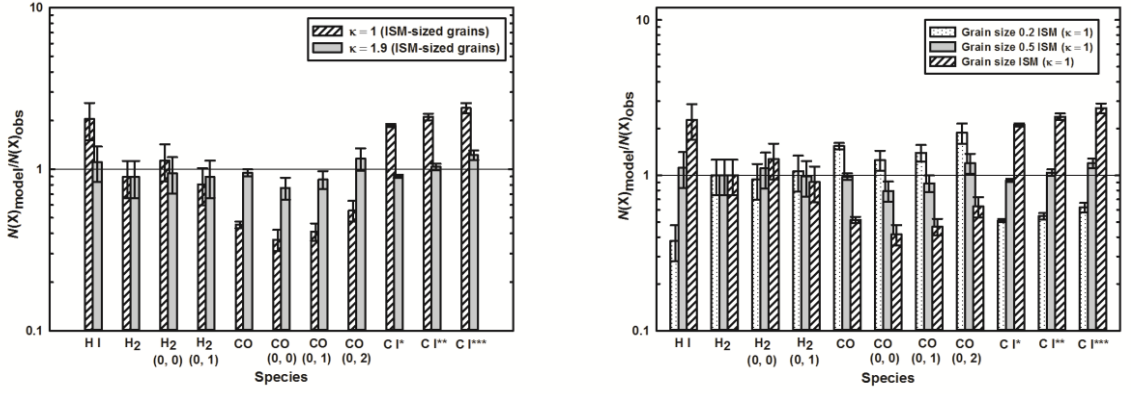

**Figure 3.** The plot shows the ratio of model-to-observed column densities for H I, H$_2$, CO and C I for the DLA at $z_{abs}$ = 2.41837 towards the quasar SDSS J143912.04+111740.5. In the left panel, the dust abundance is varied for ISM-sized grains while all other input parameters are held fixed. In the right panel, only the grain size is varied for a fixed dust abundance equivalent to the ISM abundance, to show its effect on the column densities of various species. The errors have been calculated by considering the ratio of our model predictions with the upper and lower limits of observed column density.

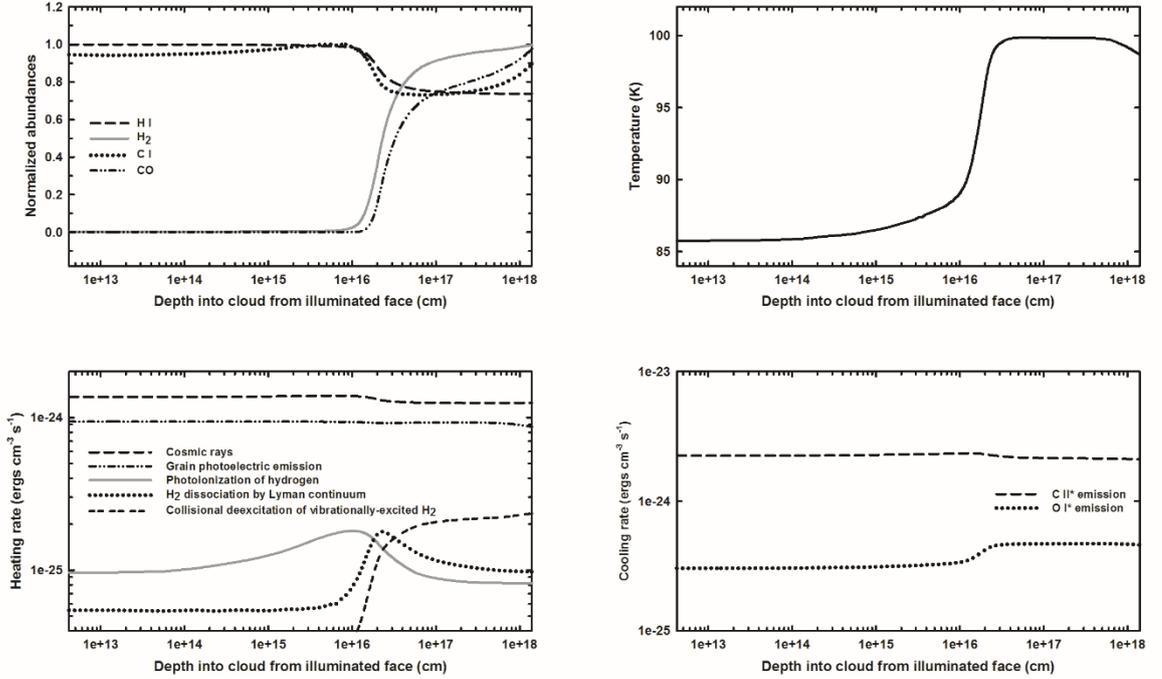

**Figure 4.** The physical conditions of the DLA at $z_{abs}$ = 2.41837 towards the quasar SDSS J143912.04+111740.5, as predicted by Model 2 are summarized in this figure. The upper left panel shows the abundance profile for H I, H$_2$, C I and CO; while the upper right panel shows the temperature variation through the DLA. The lower left and right panels show the variation of the heating caused due to cosmic rays, grain photoelectric emission, photoionization of hydrogen, H$_2$ dissociation by Lyman continuum and collisional deexcitation of vibrationally-excited H$_2$; and the cooling due to C II* and O I* emission, respectively. Model 1 shows similar variation trends.

### 2.3. DLA at $z_{abs}$ = 2.3377 towards the quasar LBQS 1232+0815

This DLA was observed using the UVES at the VLT, and the results are reported in Srianand et al. (2000), Srianand et al. (2005), Ivanchik et al. (2010) and Balashev et al. (2011). This DLA which shows simple single-component absorption, does not completely cover the background quasar, and has a covering fraction of 0.94. The observed column densities reported by Balashev et al. (2011) have been obtained after taking care of this partial coverage. We use these corrected column densities to constrain our model. Our model is well constrained by the column densities of the C I fine structure lines and the first six rotational levels of the ground vibrational state of H$_2$.

We model the system as a constant-pressure cloud with radiation incident on it from both sides. For this DLA too, we find that a blackbody radiation field of temperature 30,000-40,000 K does not reproduce the observed $H_2$ level population satisfactorily. We try the two scenarios which involve modelling only the cold neutral medium, and including the surrounding ionized region along with the cold phase, respectively. In the first scenario, we are unable to constrain $N$ (H I) and the higher rotational levels of $H_2$ simultaneously, though we try different values of radiation field intensity and cosmic ray ionization rate. So, we also explore the possibility of the DLA being situated very close to the star and hence being subject to very harsh radiation. But this scenario, as for the DLA at $z_{abs}$ = 2.41837, does not reproduce the observed mutual ratios of the C I fine structure levels. We eventually employ a hard X-ray radiation field similar to the previous DLA. The radiation extends over the energy range 1-100 keV, and we model only the cold neutral phase of the gas. We use logarithmic ionization parameter of -5.9 in the model. This radiation field has UV intensity 0.4 $G_0$ (Tielens & Hollenbach 1985). The strong X-ray radiation leads to enhanced $H_2$ population in the higher rotational levels.

The hydrogen density in our model is ~ 63 cm$^{-3}$, which is close to the range 40–60 cm$^{-3}$ derived by Srianand et al. (2005). We include cosmic rays with ionization rate of neutral hydrogen $6.3 \times 10^{-17}$ s$^{-1}$. This is lower than the Galactic background rate. This rate is constrained through the $H_2$ level population and C I lines. Cosmic rays have the most significant effect on the $H_2$ (0, 2) and (0, 3) levels. The population of the higher rotational levels in an XDR is mainly determined by the ionization parameter and density and are largely insensitive to cosmic rays. The column densities and mutual ratios of the C I lines and C II* too, are significantly affected by cosmic rays. The mean metallicity is set to 0.056 $Z_\odot$ from observation. In addition, we use observed elemental abundances for individual species from table 3 of Balashev et al. (2011). There is no observational constraint on the C abundance. However, we need to vary it to match the observed column densities of C I fine structure lines. Further, we ensure that the CO column density is lower than the non-detection limit imposed by Balashev et al. (2011). The Doppler parameter for our model is 4 km s$^{-1}$, which is close to the values reported by Ivanchik et al. (2010) (4.5 km s$^{-1}$ for the $H_2$ rotational levels J = 2 and 3, and > 3.3 km s$^{-1}$ for the J = 4 level).

This DLA has observed column densities of Fe II, Si II and S II. As we did for the DLA at $z_{abs}$ = 2.41837, we compute the dust abundance by considering S and Si individually as the undepleted species with Fe as the depleted species using Eq. (1). The dust-to-gas ratio from [Fe/S] is 0.05±0.01 and from [Fe/Si] is 0.03±0.01. For this DLA, we find that using ISM-sized dust grains with abundance 0.04 ISM (Model 1) effectively reproduces all the observed column densities. This agrees with the lower and upper limits respectively of the dust abundance from [Fe/S] and [Fe/Si]. Further, we try to incorporate smaller grains in our model and use half the dust abundance. We find that using grains of 0.5 ISM size with abundance 0.02 ISM (Model 2) replicates all the observed column densities well. The dust-to-gas ratio 0.02 ISM is also in agreement with the lower limit of the value calculated from [Fe/Si]. All other input parameters are the same for Models 1 and 2. The grains in both models follow the power law of the MRN size distribution. Since Si is known to deplete on grain surface, it is likely that Model 1 is more accurate. Interestingly, Hirashita & Ferrara (2005) have suggested the possibility of smaller grains along this sightline. However, they consider only UV interstellar radiation field in their calculations, while our model incorporated X-ray radiation as well. Besides, their calculations involve only $H_2$, and they compute a hydrogen density ≥ 2000 cm$^{-3}$ using the non-thermalized (0, 4) and (0, 5) levels of $H_2$. We successfully model both $H_2$ and C I, and determine a hydrogen density closer to the value estimated by Srianand et al. (2005) from the observations of the C I lines. However, this further stresses the need to accurately constrain the dust-to-gas ratio from the observed column densities, in order to robustly predict the grain sizes in the DLA. The input parameters of both our models are listed in Table 5, and the observed and predicted column densities of various species are in Table 6.

We plot Fig. 5 for this DLA along the lines of Figs. 1 and 3. The left panel shows the effect of using different abundance of ISM-sized grains on column densities of different species. The right panel shows the predictions of three models utilizing different grain sizes (0.2 ISM, 0.5 ISM and ISM size) where the dust-to-gas ratio is fixed at the dust abundance 0.02 ISM. All models follow the power law of the MRN size distribution. Other parameters apart from the mentioned grain properties are held fixed in the models.

Fig. 6 shows the variation of abundances, temperature, heating and cooling mechanisms with depth from one of the two illuminated faces of the cloud. The other half of the cloud will have a symmetrical profile for each of these quantities. The cloud extends to 5.41 pc, which agrees with the observationally estimated size of the neutral phase of the cloud $8.2^{+6.5}_{-4.1}$ pc (Balashev et al. 2011). The predicted gas pressure from our model for this DLA lies in the range 3330–3770 K cm$^{-3}$. The electron temperature reaches 59 K in the interior of the cloud, which is in agreement with the kinetic temperature $T_{01}$ = 67±11 K derived from the observed column densities of the rotational levels $H_2$ (0, 0) and (0, 1) (Ivanchik et al. 2010). All these values are from the predictions of Model 2. Model 1 yields similar values. The dominant heating process in this system is through cosmic rays. This accounts for 45-60 percent of the heating at various depths into the cloud. The next major contribution comes from grain photoelectric emission, photoionization of hydrogen and collisional deexcitation of vibrationally-excited $H_2$, all of which contribute 12-20 percent of the heating fraction at different depths into the cloud. While grain photoelectric heating and cosmic ray heating have almost uniform effect throughout, heating due to photoionization of hydrogen peaks at a depth ~ $10^{16}$ cm, while collisional deexcitation of $H_2$ acquires significance in the molecular regime beyond distance ~ $10^{18}$ cm and is mainly responsible for the sudden rise in temperature deep into the cloud. Cooling occurs chiefly through the C II* 158 μm and O I* 69 μm fine structure lines. C II* contributes ≥ 95 percent of the cooling fraction, with O I* responsible for a maximum of 3 percent. The heating and cooling fractions discussed here are with respect to Model 2, which uses smaller grains. Model 1 agrees to within 10 percent for most quantities. Grain photoelectric heating is higher in case of Model 1. It shows similar variation across the cloud, and agrees with Model 2 to within 20 percent. We have shown in Fig. 6, the results from Model 2.

Apart from the species we have used to constrain our model, no other species in the CLOUDY chemistry network has predicted log column density > 12.50 for this DLA.

**Table 5.** Input parameters for our CLOUDY models for the DLA at $z_{abs}$ = 2.3377 towards the quasar LBQS 1232+0815

| Parameter | Values from observation | Model 1 | Model 2 |
|---|---|---|---|
| Hydrogen density, $n_H$ | 40–60 cm$^{-3}$ [a,b] | 62-64 cm$^{-3}$ | 62-64 cm$^{-3}$ |
| Logarithmic ionization parameter | | -5.9 | -5.9 |
| [C/H] | | -1.65 | -1.65 |
| [N/H] | -2.23±0.24 [c] | -2.21 (-2.21) [d] | -2.21 (-2.21) [d] |
| [Mg/H] | -1.16±0.25 [c] | -1.25 (-1.2) [d] | -1.25 (-1.2) [d] |
| [Si/H] | -1.42±0.09 [c] | -1.47 (-1.5) [d] | -1.47 (-1.5) [d] |
| [P/H] | -1.54±0.25 [c] | -1.50 (-1.55) [d] | -1.50 (-1.55) [d] |
| [S/H] | -1.32±0.12 [c] | -1.25 (-1.32) [d] | -1.25 (-1.32) [d] |
| [Cl/H] | -1.23±0.16 [c] | -1.31 (-1.07) [d] | -1.31 (-1.07) [d] |
| [Ar/H] | -1.63±0.24 [c] | -1.48 (-1.63) [d] | -1.48 (-1.63) [d] |
| [Mn/H] | -2.22±0.11 [c] | -2.25 (-2.32) [d] | -2.25 (-2.32) [d] |
| [Fe/H] | -1.97±0.11 [c] | -2.08 (-2.05) [d] | -2.08 (-2.05) [d] |
| [Ni/H] | -2.35±0.09 [c] | -2.41 (-2.41) [d] | -2.41 (-2.41) [d] |
| Metallicity (abundance for other species) | 0.056 $Z_\odot$ [e] | 0.056 $Z_\odot$ | 0.056 $Z_\odot$ |
| Grain size range [f] | | 0.005–0.250 μm (ISM) | 0.0025–0.125 μm (0.5 ISM) |
| Dust-to-gas ratio, $\kappa$ | 0.05±0.01 (From [Fe/S])  0.03±0.01 (From [Fe/Si]) | 0.04 ISM | 0.02 ISM |
| Doppler parameter, $b$ | 4.5 km s$^{-1}$, > 3.3 km s$^{-1}$ [g] | 4 km s$^{-1}$ | 4 km s$^{-1}$ |
| Cosmic ray ionization rate, $\zeta_H$ | | 6.3 × 10$^{-17}$ s$^{-1}$ | 6.3 × 10$^{-17}$ s$^{-1}$ |

[a] Srianand et al. (2005).
[b] Gas density is determined through the rate equations of the C I fine structure levels, assuming the gas temperature to be constrained by $T_{01}$, here 67±11 K.
[c] Values from table 3 of Balashev et al. (2011) with respect to solar abundances in Lodders (2003).
[d] Abundances are with respect to the solar abundances in Grevesse et al. (2010), which is the solar standard used by CLOUDY. The abundance values in brackets are the values with respect to solar abundances in Lodders (2003).
[e] Mean metallicity is derived from the sulphur abundance (Balashev et al. 2011).
[f] MRN or MRN-like power-law size distribution.
[g] Ivanchik et al. (2010).

**Table 6.** Observed column densities and our CLOUDY model predictions for the DLA at $z_{abs}$ = 2.3377 towards the quasar LBQS 1232+0815. Model 1 uses ISM-sized grains with abundance 0.04 ISM (which agrees with the dust-to-gas ratio calculated from both [Fe/S] and [Fe/Si]), and Model 2 uses smaller-sized grains (0.5 times ISM) with abundance 0.02 ISM (which agrees with the dust-to-gas ratio calculated from [Fe/Si]).

| Species (X) | Observed column densities log $N(X)$ cm$^{-2}$ | Model 1 predicted column densities log $N(X)$ cm$^{-2}$ | Model 2 predicted column densities log $N(X)$ cm$^{-2}$ |
|---|---|---|---|
| H I | 20.90±0.08[a] | 20.99 | 20.99 |
| H$_2$ | 19.68$^{+0.08}_{-0.10}$[a] | 19.58 | 19.58 |
| H$_2$ (0, 0) | 19.45±0.10[a] | 19.40 | 19.40 |
| H$_2$ (0, 1) | 19.29±0.15[a] | 19.11 | 19.11 |
| H$_2$ (0, 2) | 16.78±0.24[a] | 16.78 | 16.78 |
| H$_2$ (0, 3) | 16.36±0.10[a] | 15.98 | 15.98 |
| H$_2$ (0, 4) | 14.70±0.06[a] | 14.84 | 14.84 |
| H$_2$ (0, 5) | 14.36±0.07[a] | 14.57 | 14.57 |
| C I* | 13.87±0.05[b] | 13.85 | 13.85 |
| C I** | 13.56±0.04[b] | 13.56 | 13.56 |
| C I*** | 12.82±0.07[b] | 12.69 | 12.69 |
| C II | 15.60±0.30[c] | 15.80 | 15.80 |
| C II* | 13.92±0.15[c] | 13.72 | 13.72 |
| N I | 14.54±0.22[d] | 14.64 | 14.64 |
| Mg II | 15.33±0.24[d] | 15.36 | 15.36 |
| Si II | 15.06±0.05[d] | 15.06 | 15.06 |
| P II | 12.86±0.24[d] | 12.94 | 12.94 |
| S II | 14.81±0.09[d] | 14.89 | 14.89 |
| Cl I | 12.97±0.14[d] | 13.10 | 13.10 |
| Ar I | 13.86±0.22[d] | 13.94 | 13.94 |
| Mn II | 12.22±0.08[d] | 12.20 | 12.20 |
| Fe II | 14.44±0.08[d] | 14.45 | 14.45 |
| Ni II | 12.81±0.04[d] | 12.83 | 12.83 |

[a] From Ivanchik et al. (2010). H$_2$ rotational level population details in table 1 therein.
[b] Column densities obtained by profile fitting and listed in table 1 of Balashev et al. (2011).
[c] Section 4.3.3 of Balashev et al. (2011).
[d] From table 3 of Balashev et al. (2011).

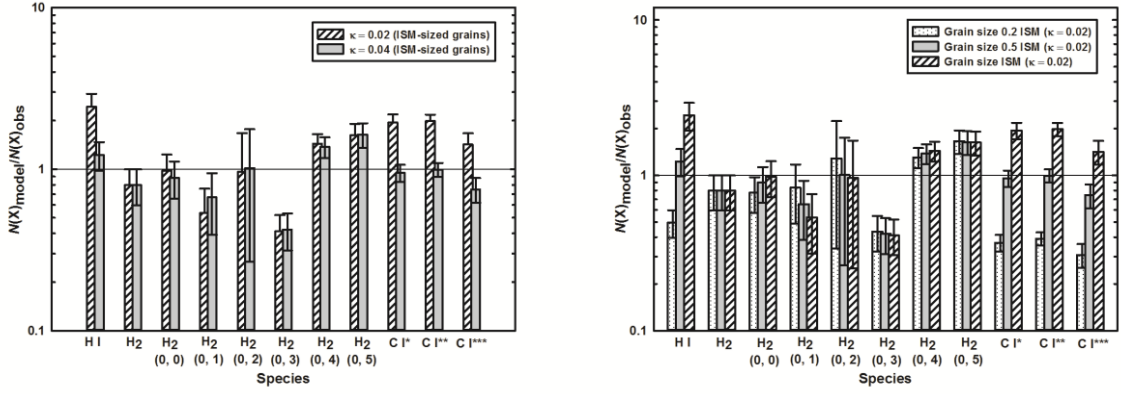

**Figure 5.** The plot shows the ratio of model-to-observed column densities for H I, C I and $H_2$ for the DLA at $z_{abs} = 2.3377$ towards the quasar LBQS 1232+0815. In the left panel, the dust abundance is varied for ISM-sized grains while all other input parameters are held fixed. In the right panel, only the grain size is varied for a fixed dust abundance of 0.02 ISM to show its effect on the column densities of various species. The errors have been calculated by considering the ratio of our model predictions with the upper and lower limits of observed column density.

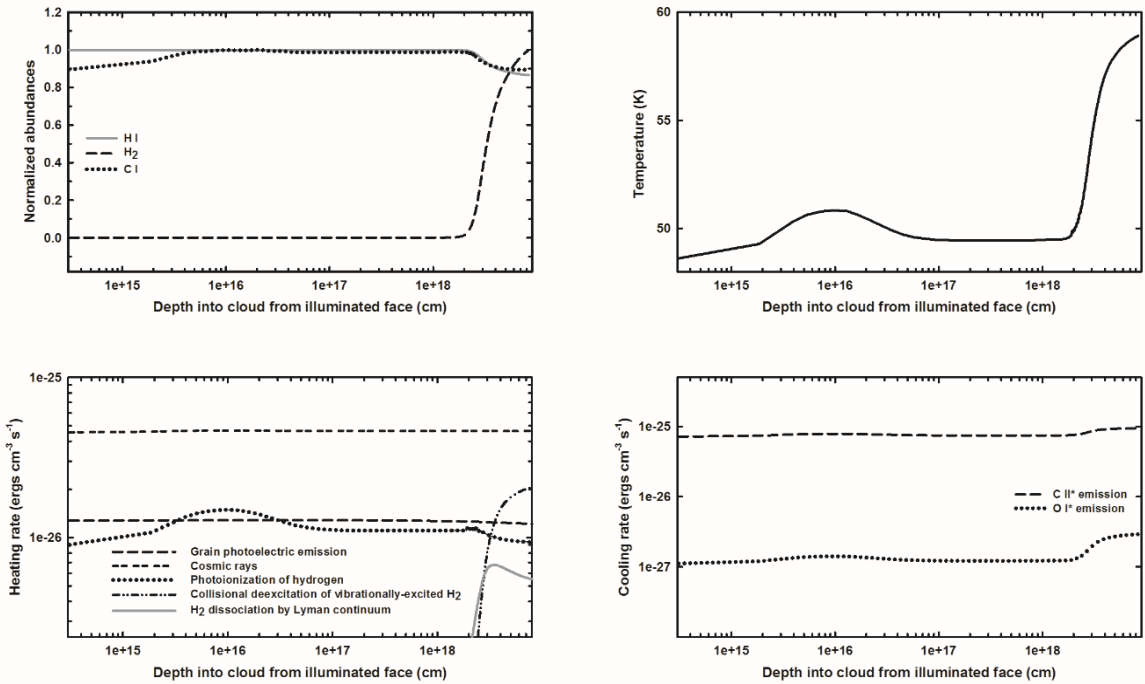

**Figure 6.** The physical conditions of the DLA at $z_{abs} = 2.3377$ towards the quasar LBQS 1232+0815, as predicted by Model 2 are summarized in this figure. The upper left panel shows the abundance profile for H I, $H_2$ and C I; while the upper right panel shows the temperature variation through the DLA. The lower left and right panels show the variation of the heating caused due to grain photoelectric emission, cosmic rays, photoionization of hydrogen, $H_2$ dissociation and collisional deexcitation of vibrationally-excited $H_2$; and the cooling due to C II* and O I* emission, respectively. Model 1 shows similar variation trends.

# 3 DISCUSSION

## 3.1. Implications for grains

Our models for the 3 DLAs show strong dependence on both dust abundance and sizes. Depending on the species that are used to calculate the dust abundance to constrain our models, we need either ISM-sized grains or smaller grains (with radius range 0.0025-0.125 μm) to successfully reproduce all the observed column densities.

Smaller grains for a given grain abundance provide more surface area, which enhances the formation of $H_2$. Thus, the observed $H_2$ column density is produced within a smaller depth into the cloud. Naturally, the amount of CO produced is also higher. The corresponding $N$ (H I) is therefore, lower. Grain size also noticeably affects the $H_2$ population in levels (0, 0), (0, 1) and (0, 2). This is because smaller grains cause more photoelectric heating and lead to an increase in temperature. The ratios of the C I fine structure lines are affected too, with the upper levels being populated more. The ratios of CO rotational level population however, are mainly influenced by CMB. They are largely unaffected by grain size and local excitation processes. We have introduced smaller grains in our models by changing the range of grain sizes, but keeping the exponent of the power-law size distribution fixed at -3.5. Alternately, a similar population of smaller grains can be also generated by retaining the ISM size range of grains but decreasing the index of the power law to -4. There are no major differences in the column densities or physical conditions predicted by the two models. However, similar effect is also produced by introducing a population of ISM-sized grains following the standard MRN power-law distribution, with twice the abundance of the smaller grains. It thus appears that our DLA models only point to the need for increased grain surface area.

Observationally, the dust extinction curves of high-$z$ DLAs are found to be similar to the Magellanic Clouds rather than the Milky Way (Ellison, Hall & Lira 2005; Khare et al. 2012; Ledoux et al. 2015). Rodrigues et al. (1997) carried out polarization studies of the interstellar dust in the Small Magellanic Cloud (SMC) and concluded that it could harbour smaller dust grains than the Milky Way. They suggest that the SMC may be rich in grains with size smaller than 0.05 μm. This also hints towards the possibility of grain sizes in high-$z$ DLAs being smaller than the Milky Way grains. Besides, other dust models using log-normal size distribution of grains and also consisting of a population of small carbonaceous grains (including PAHs) have been proposed to explain the observed infrared and microwave emission from the diffuse interstellar medium of the Milky Way and the Magellanic Clouds (Weingartner & Draine 2001; Li & Draine 2001; Draine & Li 2007). We did not include PAHs for the DLA models we have presented here as PAHs have not yet been detected in DLAs.

Throughout the above calculations, it has been assumed that grains are perfectly spherical and homogenous in structure and composition. However, the actual structure of dust grains is unlikely to be so simple. Mathis & Whiffen (1989) proposed a grain model for composite grain particles containing voids in the structure. Taking into account this fluffy character of grains would also increase the available grain surface area for a given abundance. So, we also explore the effect of fluffy grains in our DLA models. Instead of smaller-sized grains, we now include ISM-sized amorphous carbon and silicate particles (radius range 0.005-0.250 μm), containing a specified proportion of vacuum by volume known as porosity. We find that for ISM-sized grains of porosity 0.55, the predicted column densities are very similar to the predictions of our models for all 3 DLAs. In Fig. 7 we compare three models – (a) with ISM-sized compact grains, (b) with ISM-sized fluffy grains of porosity 0.55, and (c) with 0.5 ISM-sized compact grains – all having the same dust abundance. The predicted column densities are in agreement with the column densities predicted by the models we have discussed earlier. However, this is only a rudimentary and simple approach to study the effect of introducing fluffy grains in our models. The actual structure and distribution of dust grains may be more complex. It is possible that only some dust grains are fluffy while others are compact. Besides, there is no consensus on the expected porosity in interstellar dust. Mathis & Whiffen (1989) consider high porosity of 0.8 in their grain model, while Heng & Draine (arXiv:0906.0773) through their X-ray and optical study of grain aggregates find that porosities must be ≤ 0.55.

Our model predictions are limited by the dust-to-gas ratio constrained from observations of different combinations of refractory and volatile elements. In order to make any robust conclusions about either dust grain sizes or porosity in DLAs, it is highly essential to have an accurate estimate of the dust-to-gas ratio.

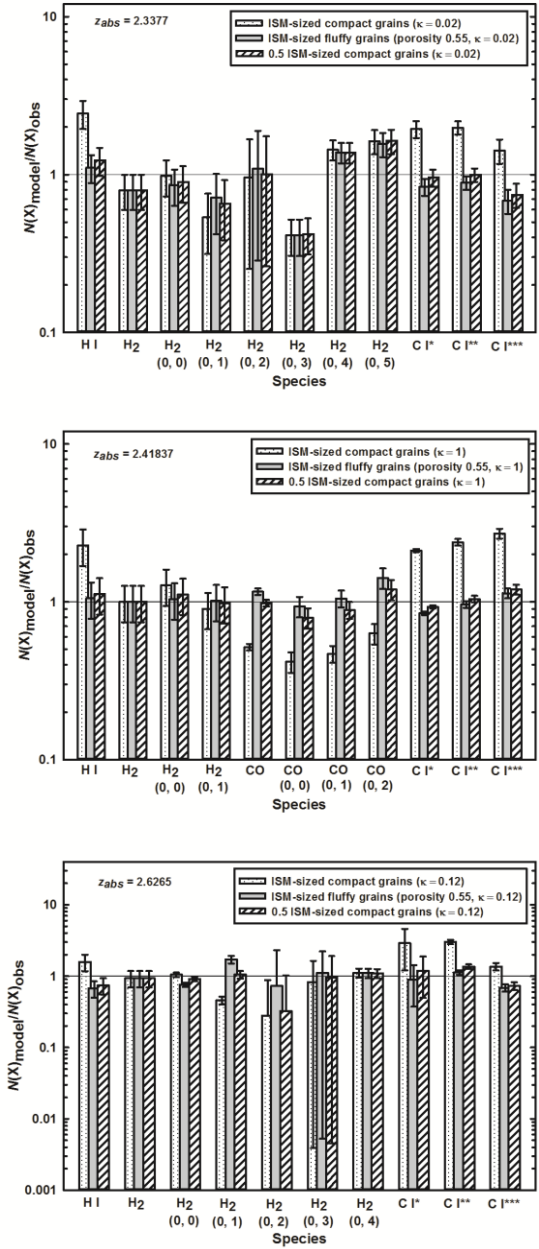

**Figure 7.** Ratio of model-to-observed column densities of H I, H$_2$ and C I for the 3 DLAs for three different models – (a), with ISM-sized compact grains following the MRN power-law distribution; (b), with ISM-sized fluffy grains with 0.55 porosity following the MRN power-law distribution; and (c), with grain sizes in the range 0.0025–0.125 μm (0.5 times the ISM) following the MRN power-law size distribution. Models (b) and (c) produce almost identical results. The upper, middle and lower panels are for the DLAs at $z_{abs}$ = 2.3377, 2.41837 and 2.6265 towards the quasars LBQS 1232+0815, SDSS J143912.04+111740.5 and FBQS J081240.6+320808 respectively.

## 3.2. Radiation field

We consider three sources of ionizing radiation in our models – Haardt-Madau metagalactic background, CMB and the radiation from local star formation. To mimic the radiation from O/B type stars, the local interstellar radiation field in DLAs is generated as a blackbody spectrum of temperature 30,000-40,000 K. For the DLA at $z_{abs}$ = 2.6265, we use such a blackbody radiation field of temperature of 40,000 K. However, for the DLAs at $z_{abs}$ = 2.3377 & 2.41837, we have to use an X-ray radiation field to reproduce all the observed quantities satisfactorily. As mentioned earlier, the X-ray radiation field in our models has spectral energy distribution as described in Maloney et al. (1996), with significant intensity in the energy interval 1-100 keV (~ 0.12-12 Å) and negligible intensity outside it. Regions in the vicinity of such X-ray radiation sources form XDRs and are different from photon-dominated regions (PDRs) which are illuminated by UV radiation (Tielens & Hollenbach 1985). In addition to grain photoelectric emission, photoionization of hydrogen and H$_2$ photodissociation through the Lyman continuum are the other important heating mechanisms in XDRs (Meijerink & Spaans 2005, Gay et al. 2012). The contribution of these two processes in our models for the DLAs at $z_{abs}$ = 2.3377 and 2.41837 can be seen in Figs. 4 & 6. The rotational level population of $H_2$ is significantly changed on account of the increased $H_2$ photodissociation (Gay et al. 2012).

In Fig. 8, as a demonstration, we plot the spectral energy distribution (SED) of all the radiation sources in our model to show the individual contribution of the metagalactic background, CMB and X-ray radiation field to the incident continuum. The SED of the metagalactic background and the CMB vary with redshift. However, they are of the same order of magnitude for the three DLAs we consider, since the redshift difference of the systems is small. The X-ray radiation field has significant flux only in the energy interval 1-100 keV with the strength of the radiation depending only on the ionization parameter and independent of redshift. Hence, we show the SED of the contributing radiation sources only for the DLA at $z_{abs}$ = 2.3377, where the X-ray radiation field has logarithmic ionization parameter of -5.9. From Fig. 8 it can be seen that the UV region of the incident radiation field is clearly dominated by the metagalactic background. The X-ray radiation dominates over the metagalactic background in the wavelength interval 0.12-12 Å. We plot a blackbody spectrum of 40,000 K which is the typical temperature expected from O-type stars and compare it with the X-ray radiation field. It is clearly evident that a UV blackbody radiation field does not produce significant flux at higher energies like the X-ray radiation field. We note that in our model, the total incident continuum generated by considering the metagalactic background, CMB and X-ray radiation field, passes through intervening gas before striking the illuminated face of the DLA. As a result, the continuum is completely attenuated in the wavelength range ~ 10-912 Å due to absorption, mainly by atomic hydrogen and helium. This is similar to absorption that takes place through gas in the Galactic plane and is necessary to be taken into account. In Fig. 8, we use dotted lines to indicate the attenuated portion of the continuum.

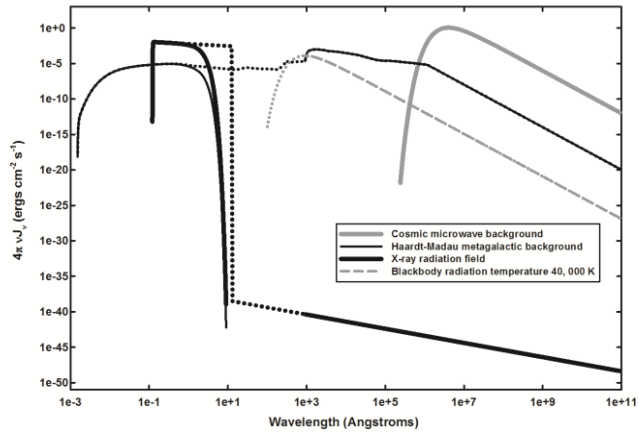

**Figure 8.** Spectral energy distribution of the radiation sources in our model for the DLA at $z_{abs}$ = 2.3377 towards the quasar LBQS 1232+0815 – Haardt-Madau metagalactic background, CMB, X-ray radiation field and a blackbody spectrum of temperature 40,000 K. The attenuated portion of the continuum is represented by dotted lines. Clearly, X-ray radiation is the dominant radiation source in the high energy interval corresponding to the wavelength range 0.12-12 Å. UV blackbody spectra do not have significant flux in this range.

We also repeat our model for the DLA at $z_{abs}$ = 2.3377 by replacing the X-ray radiation field with blackbody radiation of temperature 40,000 K and compute two models – one, with attenuation of the incident radiation field (column densities described in Table 7 as UV model 1) as discussed above, and two, without (UV model 2 in Table 7). The reference XDR model is Model 2 already discussed in section 2.3. In Table 7, we compare the predicted column densities of the 3 models thus obtained. All other parameters are retained constant, including the logarithmic ionization parameter of -5.9. UV model 1 produces more $N$ (H I), but the higher rotational levels of $H_2$ are not sufficiently excited. Higher cosmic ray ionization rate would increase the population of these levels, but would also simultaneously produce higher $N$ (H I) and $N$ (C II*). It is evident that X-ray radiation is necessary to excite the higher $H_2$ rotational levels at a lower column density of H I. As previously mentioned in sections 2.2 & 2.3, making the radiation field directly incident on the DLA without considering the intervening absorption, affects the C I fine structure levels. This is clearly evident from UV model 2. It shows $N$ (C I**)/$N$ (C I*) > 0 (log-scale), whereas observations show $N$ (C I**)/$N$ (C I*) < 0. It also severely reduces the population of the $H_2$ (0, 1) level as the harsh radiation pumps the higher rotational levels.

XDRs are well-known in the local universe (García-Burillo et al. 2010; Aalto et al. 2011). However, hardly any XDRs are known at high redshift. Recently, Gallerani et al. (2014) have suggested the possibility of a composite PDR-XDR model to explain the CO (17-16) emission line toward the quasar SDSS J114816.64+525150.3 at $z$ = 6.4. Our models clearly show that the DLAs at $z_{abs}$ = 2.3377 & 2.41837 are high-$z$ XDRs. However, it is difficult to comment on the nature of the X-ray source giving rise to the XDR. As per Maloney et al. (1996), XDRs can be formed either in regions associated with massive star formation, or in the vicinity of compact objects. The XDRs in the high-$z$ DLAs we have modelled may thus be the result of intense star formation in the DLA, or due to the presence of an active galactic nucleus (AGN) near the DLA.

**Table 7.** CLOUDY model predictions for the DLA at $z_{abs}$ = 2.3377 towards the quasar LBQS 1232+0815, when using UV radiation field instead of X-ray radiation. Models are computed with and without attenuation due to intervening absorption.

| Species (X) | Observed column densities log $N(X)$ cm$^{-2}$ | XDR model [a] log $N(X)$ cm$^{-2}$ | UV model 1 [b] log $N(X)$ cm$^{-2}$ | UV model 2 [c] log $N(X)$ cm$^{-2}$ |
|---|---|---|---|---|
| H I | 20.90±0.08[d] | 20.99 | 21.26 | 20.31 |
| H$_2$ | 19.68$^{+0.08}_{-0.10}$[d] | 19.58 | 19.58 | 19.58 |
| H$_2$ (0, 0) | 19.45±0.10[d] | 19.40 | 19.49 | 19.57 |
| H$_2$ (0, 1) | 19.29±0.15[d] | 19.11 | 18.85 | 18.02 |
| H$_2$ (0, 2) | 16.78±0.24[d] | 16.78 | 15.60 | 15.13 |
| H$_2$ (0, 3) | 16.36±0.10[d] | 15.98 | 14.59 | 14.83 |
| H$_2$ (0, 4) | 14.70±0.06[d] | 14.84 | 13.58 | 14.10 |
| H$_2$ (0, 5) | 14.36±0.07[d] | 14.57 | 13.13 | 14.11 |
| C I* | 13.87±0.05[e] | 13.85 | 13.94 | 13.69 |
| C I** | 13.56±0.04[e] | 13.56 | 13.62 | 13.73 |
| C I*** | 12.82±0.07[e] | 12.69 | 12.71 | 13.27 |
| C II | 15.60±0.30[f] | 15.80 | 16.06 | 15.20 |
| C II* | 13.92±0.15[f] | 13.72 | 13.85 | 13.84 |
| N I | 14.54±0.22[g] | 14.64 | 14.90 | 14.07 |
| Mg II | 15.33±0.24[g] | 15.36 | 15.63 | 14.77 |
| Si II | 15.06±0.05[g] | 15.06 | 15.32 | 14.48 |
| P II | 12.86±0.24[g] | 12.94 | 13.20 | 12.36 |
| S II | 14.81±0.09[g] | 14.89 | 15.15 | 14.32 |
| Cl I | 12.97±0.14[g] | 13.10 | 13.26 | 12.55 |
| Ar I | 13.86±0.22[g] | 13.94 | 14.20 | 13.37 |
| Mn II | 12.22±0.08[g] | 12.20 | 12.46 | 11.63 |
| Fe II | 14.44±0.08[g] | 14.45 | 14.71 | 13.87 |
| Ni II | 12.81±0.04[g] | 12.83 | 13.09 | 12.26 |

[a] Model 2 from Section 2.3.
[b] Blackbody radiation field of logarithmic ionization parameter of -5.9 used in the XDR model. Attenuation of the incident radiation field due to intervening absorption is considered.
[c] Blackbody radiation field of logarithmic ionization parameter of -5.9 used in the XDR model. The incident radiation field is not attenuated.
[d] From Ivanchik et al. (2010). H$_2$ rotational level population details in table 1 therein.
[e] Column densities obtained by profile fitting and listed in table 1 of Balashev et al. (2011).
[f] Section 4.3.3 of Balashev et al. (2011).
[g] From table 3 of Balashev et al. (2011).

### 3.3. Geometry and thickness of H$_2$-bearing gas

DLAs have generally been modelled as plane parallel slabs with radiation illuminating one face of the cloud. But we find that considering slab geometry with radiation incident from both sides, yields column densities which are a closer match to observations. In this section, we first compare our model for each DLA with a corresponding model in which radiation is incident only on one side.

For the DLA at $z_{abs}$ = 2.3377, replacing the two-sided radiation field with one-sided radiation does not affect the H$_2$ level population significantly. It produces ~ 0.1 dex lower $N$ (H I) and $N$ (C II*). It also reduces the C I fine structure level population by 0.1 dex, but does not affect their mutual ratios. We find that the model with two-sided radiation reproduces observation better. Merely increasing C abundance would not make the model with one-sided radiation to match the observed values since this would also adversely affect the H$_2$ level population. For the DLA at $z_{abs}$ = 2.6265, using radiation only on one face of the cloud yields $N$ (H I) and $N$ (C II*) lower by ~ 0.15 dex. We also see a similar effect on the column densities of H$_2$ (0, 2), (0, 3) and (0, 4) levels. $N$ (C I***) too falls below the range of observational error by 0.2 dex. Increasing the intensity of UV radiation to improve the higher rotational levels of H$_2$ would also disturb the H$_2$ (0, 0) level which our model matches only to the lower limit from observation. We conclude for this system too, that two-sided radiation yields a better-fitting model. For the DLA at $z_{abs}$ = 2.41837, we do not observe major differences between $N$ (H I) and the H$_2$ level population predicted by both models. A small difference of ~ 0.04 dex is seen in the predicted $N$ (CO) and C I fine structure population. We retain the double-sided radiation field for this DLA too, for the sake of consistency.

Srianand et al. (2012) study 21-cm and H$_2$ absorption in high-z DLAs and suggest that H$_2$ absorption arises from compact regions ≤ 15 pc across. Here, we compare the predicted size of the DLA clouds from our models with the limit set by Srianand et al. (2012). Our model for the DLA at $z_{abs}$ = 2.3377 predicts a size of 5.41 pc for the cold neutral phase of the DLA. Based on the position of the H I/H$_2$ transition in the upper left panel of Fig. 6, we can infer that the H$_2$-bearing region of the DLA has size < 1.5 pc. Balashev et al. (2011) have calculated the size of the H$_2$-bearing core of this DLA and the neutral envelope to be 0.15±0.05 pc and 8.2$^{+6.5}_{-4.1}$ pc respectively. The DLA at $z_{abs}$ = 2.41837 has a very small extent of 1.02 pc. Our model traces mainly the molecular region, and hence this size can be approximated to be the extent of the molecular core of the DLA. The DLA at $z_{abs}$ = 2.6265 has a size 2.95 pc, and from the H I/H$_2$ transition point in Fig. 2, the extent of the H$_2$-bearing core can be deduced to be < 0.9 pc. Jorgenson et al. (2009) set a lower limit of 0.1 pc to the size of the cloud. The sizes of these DLAs show that H$_2$ absorption arises from compact regions and are clearly in agreement with the limit set by Srianand et al. (2012).

### 3.4. Pressure

The total pressure remains constant throughout the cloud in our DLA models, but the thermal component of pressure varies at different depths into the cloud. If the quasar sightline traced both the WNM and CNM within the DLA, there would have been significant variation in temperature and density across the cloud. In the Galactic ISM, the warm and cold neutral phases are known to co-exist in pressure equilibrium (Field et al. 1969; Wolfire et al. 1995, 2003). We assumed constant pressure in our models as we did not have a priori knowledge of the temperature and density profile within the cloud. However, we find that the sightlines trace only the CNM, and temperature variation with depth is not significant in our models. Thus, in the absence of magnetic and hydrodynamic pressure, our constant pressure models are equivalent to constant density models.

Srianand et al. (2005) study the physical conditions in $H_2$-bearing high-$z$ DLAs. For a sample of 33 DLAs, they find that the gas pressure lies in the range 824-30,000 K cm$^{-3}$. 40 percent of the systems have pressure higher than 3000 K cm$^{-3}$. For the DLAs at $z_{abs}$ = 2.3377, 2.41837 and 2.6265, the thermal pressure varies in the range 3330–3770, 5620–5660 and 4150–4660 K cm$^{-3}$ respectively. Clearly, the pressure in our DLA models agrees with the higher-end of pressure seen in $H_2$-bearing DLAs.

From studies of C I fine structure excitation, it is known that thermal pressure in the diffuse cold neutral medium of our Galaxy follows a log-normal distribution with a peak at $P/k$ ~ 3800 K cm$^{-3}$ (Jenkins & Tripp 2011). The pressure in the DLAs at $z_{abs}$ = 2.41837 and 2.6265 is higher than this Galactic median pressure, while the DLA at $z_{abs}$ = 2.3377 has slightly lower pressure. Compared to these values, the typical pressure seen in giant molecular clouds in the Milky Way and in neighbouring galaxies is much higher with $P/k$ ~ $10^5$–$10^7$ K cm$^{-3}$ (Tatematsu et al. 2004). These molecular clouds are sites of intense star formation activity. Hence, DLAs with their comparatively lower pressure, are not likely to be intense star-forming regions. This agrees with the low star formation rates, $\leq$ 1–10 M$_\odot$ yr$^{-1}$ deduced for high-$z$ DLAs (Wolfe & Chen 2006; Fumagalli et al. 2010; Rahmani et al. 2010). However, as DLAs do harbour molecular gas, it is possible that they may eventually evolve, leading to the development of an intense star formation region (Jorgenson et al. 2009).

## 4 CONCLUSIONS

We have performed detailed numerical simulation of 3 DLAs: at $z_{abs}$ = 2.3377 towards LBQS 1232 + 082, at $z_{abs}$ = 2.41837 towards SDSS J143912.04+111740.5 and at $z_{abs}$ = 2.6265 towards FBQS J081240.6+320808 using the spectral synthesis code CLOUDY. We have reproduced most of the observed column densities satisfactorily. Our conclusions are as follows:

- The above three DLAs are well-reproduced constant-pressure clouds in the radiation field due to local star formation. The radiation field for the DLA at $z_{abs}$ = 2.6265 is similar to a blackbody spectrum of temperature $4\times10^4$ K and intensity $5 \times 10^{-4}$ ergs cm$^{-3}$ s$^{-1}$. The DLAs at $z_{abs}$ = 2.3377 and 2.41837, are irradiated by hard X-ray radiation and are similar to X-ray Dominated Regions. The logarithmic ionization parameters are -5.9 and -5 for the respective systems. In all the systems, we find that the radiation field illuminates the DLA from both sides.
- Density lies in the range 62-64, 60–61 and 89–94 cm$^{-3}$ for the DLAs at $z_{abs}$ = 2.3377, 2.41837 and 2.6265 respectively, while the corresponding electron temperature reaches 59, 99 and 48 K in the interior of the respective DLAs. The predicted gas pressure in our 3 DLA models agrees with the pressure observed in $H_2$-bearing DLAs and is lower than the pressure characteristic of intense star-forming regions in giant molecular clouds.
- The cosmic ray ionization rates are $6.3 \times 10^{-17}$, $1.4 \times 10^{-15}$ and $2 \times 10^{-17}$ s$^{-1}$ for the DLAs at $z_{abs}$ = 2.3377, 2.41837 and 2.6265 respectively, spanning a range ~ 2 dex.
- The dust-to-gas ratio constrained from observed column densities of undepleted species such as Si, S and Zn yield different values. Depending on the dust abundance constraint imposed on our models, we find that the DLAs could either have ISM-sized grains or smaller grains (0.5 times the ISM). Alternately, we also replicate the results by using fluffy ISM-sized grains of porosity 0.55. All this points to a need for increased grain surface area. Robust conclusions about the dust grain size or porosity can only be drawn by imposing accurate constraints on the dust-to-gas ratio.
- We study the main heating and cooling processes in the 3 DLAs. The dominant heating processes vary in the 3 systems. The DLA at $z_{abs}$ = 2.3377 is dominated by cosmic ray heating, and collisional deexcitation of vibrationally-excited $H_2$ is the second important heating mechanism in the molecular region. Heating in the DLA at $z_{abs}$ = 2.41837 is mainly provided by cosmic rays with significant contribution from grain photoelectric emission. Here too, collisional deexcitation of vibrationally-excited $H_2$ is an important heating mechanism in the molecular region. For the DLA at $z_{abs}$ = 2.6265, grain photoelectric heating is the most significant process. Cooling in all 3 systems occurs mainly through C II* 158 μm line emission.
- The sizes of the DLAs at $z_{abs}$ = 2.3377, 2.41837 and 2.6265 are 5.41 pc, 1.02 pc and 2.95 pc respectively, and in agreement with the idea that $H_2$ absorption arises from compact regions which are less than 15 pc across.
- We also present our predictions for species whose transitions may lie outside the wavelength range of observations, but whose log column densities are higher than 12.50 (cm$^{-2}$). We predict column densities of C II*, OH, OH$^+$ and $H_2O$ for the DLA at $z_{abs}$ = 2.41837, and OH and HCl for the DLA at $z_{abs}$ = 2.6265.

**ACKNOWLEDGEMENTS**

This work is funded by DST. Gargi Shaw acknowledges the DST project (D.O. No. SR/FTP/PS-133/2011). The authors would like to thank the referee for his/her valuable comments. We also thank Mohit Pandey for his help.